\documentclass[print,sigconf, authorversion]{acmart}

\usepackage{dcolumn}
\usepackage{tabularx}
\newcolumntype{R}{>{\footnotesize}X}
\usepackage{quoting}
\quotingsetup{font=itshape,vskip=2pt, leftmargin=5pt}

\AtBeginDocument{%
  \providecommand\BibTeX{{%
    \normalfont B\kern-0.5em{\scshape i\kern-0.25em b}\kern-0.8em\TeX}}}

\copyrightyear{2020} 
\acmYear{2020} 
\setcopyright{acmcopyright}\acmConference[NordiCHI '20]{Proceedings of the 11th Nordic Conference on Human-Computer Interaction: Shaping Experiences, Shaping Society}{October 25--29, 2020}{Tallinn, Estonia}
\acmBooktitle{Proceedings of the 11th Nordic Conference on Human-Computer Interaction: Shaping Experiences, Shaping Society (NordiCHI '20), October 25--29, 2020, Tallinn, Estonia}
\acmPrice{15.00}
\acmDOI{10.1145/3419249.3420134}
\acmISBN{978-1-4503-7579-5/20/10}

\begin{document}

\title{Embracing Companion Technologies}


\author{Jasmin Niess}
\orcid{0000-0003-3529-0653}
\affiliation{
\institution{University of Bremen}
\city{Bremen}
\country{Germany}}
\email{niessj@uni-bremen.de}

\author{Pawe\l{} W. Wo\'{z}niak}
\orcid{0000-0003-3670-1813}
\affiliation{
\institution{Utrecht University}
\city{Utrecht}
\country{the Netherlands}}
\email{p.w.wozniak@uu.nl}

\renewcommand{\shortauthors}{Niess and Wo\'{z}niak}

\begin{abstract}
As an increasing number of interactive devices offer human-like assistance, there is a growing need to understand our experience of interactive agents. When interactive artefacts become intertwined in our everyday experience, we need to make sure that they assume the right roles and contribute to our wellbeing. In this theoretical exploration, we propose a reframing of our understanding of the experience of interactions with everyday technologies by proposing the metaphor of \emph{companion technologies}. We employ theory in the philosophy of empathy to propose a framework for understanding how users develop relationships with digital agents. The experiential framework for companion technologies provides connections between the users' psychological needs and companion features of interactive systems. Our work provides a theoretical basis for rethinking the user experience of everyday artefacts with an empathy-oriented mindset and poses future challenges for HCI.
\end{abstract}



\begin{teaserfigure}
     \centering
     \includegraphics[height=4cm]{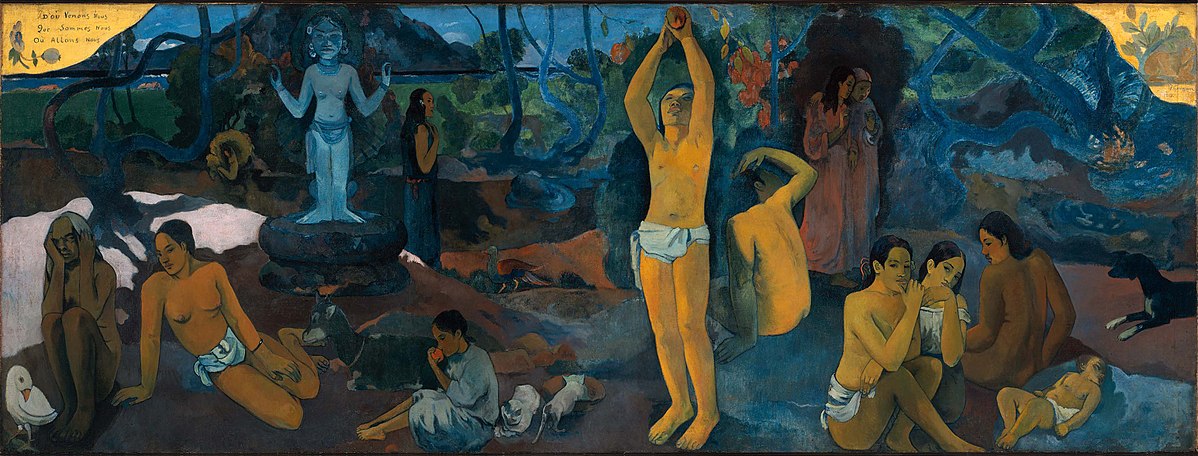}
     \caption{\emph{D'o\`{u} Venons Nous / Que Sommes Nous / O\`{u} Allons Nous} (Where do we come from? What are we? Where are we going?) by Paul Gauguin. Museum of Fine Arts, Boston.}\label{fig:gauguin}
\end{teaserfigure}
\maketitle

\section{Introduction}
\begin{quoting}
    In a sense, my smartphone has been my longest relationship.
\end{quoting}

Feeling connected to other beings and having positive interpersonal relationships makes us happy~\cite{lucas2006does}. An essential aspect of satisfactory human relationships is empathy~\cite{davis2018empathy,cramer2010perceived}. It has the power to bridge the divide between two entities, bring them closer together and support long-lasting relationships. However, humans can experience not only a bond between themselves and other human beings, but also, amongst others, between themselves and pets or even inanimate objects.

Concurrently, developing emotional relationships with inanimate objects is a strong motif in popular culture. Aaron Chervenak, the author of the epigraph above, famously married his smartphone\footnote{https://www.news24.com/You/Archive/this-man-married-his-smartphone-for-a-very-good-reason-20170728}. In the film \emph{Her} by Spike Jonze~\cite{her_2013}, a man falls in love with the artificially intelligent virtual assistant Samantha. The video game \emph{Detroit: Become Human}~\cite{epic} puts the player in the middle of an android revolution where the boundary between human and robot emotion is blurred. Yet, life with (or next to) artificial beings is not only an inspiring topic for artists, but it increasingly becomes part of our experience of the world. 

Beyond futuristic visions, interactive technologies which offer human-like assistance are becoming ubiquitous. With many of us living with a smart speaker, and some reprimanding their fellow users for not being kind to the device~\cite{Sciuto:2018:HAW:3196709.3196772}, digital beings affect our everyday interactions with the world and become part of our emotions, evoking \emph{empathy}. Consequently, empathetic and affectionate feelings towards objects have become a topic present in a variety of research areas~\cite{belk1988possessions}. For instance, Human-Robot Interaction (HRI) researchers pursue the idea of robotic agents that can evoke emotion and eventually become friends~\cite{Paiva2017}. Marketing theory explores ways to exploit the potential of bonds between users and specific brands or products to increase sales~\cite{thomson2005ties}. Philosophers have been discussing our relationship with inanimate artefacts for more than a hundred years~\cite{lipps1906einfuhlung}. Such a multitude of perspectives on our future emotional relationships with technology poses a challenge to Human-Computer Interaction (HCI). 

In HCI, a number of works reported on users developing relationships with interactive artefacts (e.g.~\cite{Sciuto:2018:HAW:3196709.3196772,montalvan2017adaptation,Sung:2007:MRR:1771592.1771601,Purington:2017:AMN}). However, these accounts are often focused on one design or one specific system. For a long time, HCI has also aimed to understand the complexity of the user experience of technologies on the meta level. Hassenzahl~\cite{hassenzahl2003thing} classified the experience into pragmatic and hedonic in his seminal work. Later, the inclusion of eudaimonic qualities was postulated by Mekler and Hornb{\ae}k~\cite{Mekler:2016:MPL:2858036.2858225}. While these works offer a comprehensive view of user experience, they do not satisfyingly address the ways in which artefacts can fulfil social needs or the development of feelings towards objects. In these works, technology mostly assumes one of the two roles: it either mediates empathy between people, or it provides a feeling of being understood by the technology. The understanding of reciprocal empathy between humans and technology, feeling what the object is experiencing and vice versa is currently missing in models of user experience. This points to a lack of a theoretical understanding of empathy in HCI.

To mitigate this, this essay uses work on empathy for objects from philosophy and accounts of user needs from past literature in HCI and psychology to ground a discussion around an alternative conceptualisation of user experience that goes beyond existing understandings~(e.g.~\cite{hassenzahl2003thing,Mekler:2016:MPL:2858036.2858225}). We introduce the concepts of \emph{companion technologies}, i.e. interactive artefacts that evoke empathy.

In line with philosophical methodology, the idea for this paper started with a sense of wonder mixed with curiosity. We reflected about notions of experiential qualities with interactive artefacts in HCI. Having conducted a study of Amazon Alexa users, one of the authors of this paper realised that they usually greeted their robot vacuum cleaner when returning from work. They also noticed that they were not becoming friends with the Pepper robot in their office, despite the robot's best efforts. Consequently, we started to formulate questions about the role of technology in our lives and slowly discovered that we had difficulties describing these phenomena with the tools, theories and frameworks we already had at hand. We asked: what does it mean to get used to an interactive artefact? When do objects become part of our lives? Did we develop a relationship with technologies? To answer this, we decided to go back to basics: fine arts, pop culture and philosophy. As an homage to this intellectual journey and because it served as a helpful framework for our discussion, the structure of this paper is inspired by Gauguin's painting \emph{D'o\`{u} Venons Nous / Que Sommes Nous / O\`{u} Allons Nous} (Where do we come from? What are we? Where are we going?), also displayed in Figure \ref{fig:gauguin}.

Based on our theoretical exploration, we derive a framework for companion technologies, which focuses on understanding the experiential aspects of a socially-embedded technology. The framework introduces a new dimension of user experience---empathetic experience---which is characterised by four concepts: \textit{minded}, \textit{feeling-experience}, \textit{reflective} and \textit{social significance}. The purpose of the framework is to empower designers and HCI scholars alike to think about future interactive systems as companion technologies, anticipate user empathy for the designed artefacts and assure that companion technologies seamlessly integrate with everyday experience. We use the term `framework' to describe the construct proposed in this work to highlight that we describe a system of ideas and beliefs that can help make design decisions. Further, we hint at Edelson's~\cite{edelson2002design} concept of frameworks as we aim to name the set of qualities that artefacts must possess to be \emph{companion technologies}.




Our inquiry explores the premise that while companion technologies are relatively novel in our society, empathy for everyday objects is an established theme in the humanities. In the remainder of this paper, we first outline HCI work with a focus on concepts of user experience. We then relate the existing theoretical understanding of user experience to the theory of psychological needs. Next, we critically discuss selected works from HCI focusing on artificial agents, reviewing our current understanding of users' relationships with objects. This is followed by an overview of philosophical views on empathy for objects. These theoretical foundations contribute to our framework for companion technologies. We then present the framework, detailing its components. We use two examples from past research to show how the lens of companion technologies can help understand the users' empathic relationships with interactive technologies. We further show how companion technologies complement other views of socially embedded technologies and discuss the limitations of our approach.  Our work includes future challenges for HCI that emerge from framing interactive artefacts as companion technologies.





\section{Where Do We Come From?: Understanding the User Experience}
HCI has an established history of recognising the importance of experiential qualities in interactive technologies and embedding them in design. Past HCI research on experiential qualities identified pragmatic~\cite{hassenzahl2003thing}, hedonic~\cite{hassenzahl2003thing} and eudaimonic~\cite{Mekler:2016:MPL:2858036.2858225} experiential aspects. HCI research on experiential qualities builds on and is inspired by a variety of different research traditions (e.g. philosophy, psychology). This leads to some conceptual similarities to other fields as well as some differences. To explore the notion of companion technologies, we engaged with the current understanding of user experience in HCI as well as with research traditions such as positive psychology that define some of the experiential concepts (e.g. eudaimonia~\cite{huta2014eudaimonia}) somewhat differently.

Almost two decades ago, Hassenzahl~\cite{hassenzahl2003thing} discussed pragmatic and hedonic product qualities. In his work, he emphasises that the perceived character of a product leads to emotional consequences, such as the experience of joy. Pragmatic, instrumental product qualities encompass aspects such as utility, usability, efficiency and usefulness. Hedonic, non-instrumental product characteristics can encompass aspects such as aesthetic appeal, fun, stimulation and joy~\cite{Diefenbach:2014:HHI:2598510.2598549}.  
Mekler and Hornb{\ae}k~\cite{Mekler:2016:MPL:2858036.2858225} extended the notion of user experience and introduced eudaimonic quality to the HCI community. In their empirical inquiry, they found that hedonia was mostly about ephemeral moments of pleasure, whereas eudaimonic qualities were characterised through striving towards personal growth, a focus on self-development and personal goals. These findings are in line with insights from psychological works on hedonia and eudaimonia~\cite{huta2014eudaimonia}.
Interestingly, in Mekler and Hornb{\ae}k's work~\cite{Mekler:2016:MPL:2858036.2858225}, the hedonic and the eudaimonic were not strongly correlated to the need for relatedness. This suggests that the experience of meaning and the fulfilment of the human need for relatedness might be addressed through experiences that are not necessarily driven by hedonic or eudaimonic motives.

Later, Mekler and Hornb{\ae}k~\cite{Mekler:2019:FEM:3290605.3300455} extended their previous work by deriving a framework of the experience of meaning in HCI. Their framework includes five components: connectedness, purpose, coherence, resonance and significance. Connectedness can be described as feeling connected to oneself and the world. Previous experiences as well as interactions with the world are connected to what is happening right now in the present moment, and shape the experience of meaning. Feeling connected to oneself is the basis of experience of meaning. However, Mekler and Hornb{\ae}k emphasised that, due to its elusive nature, connectedness is the most difficult component of their framework to design for. Our work is aimed to explore this concept further. 

While these theoretical approaches to experience encompass a wide spectrum of the ways interactive artefacts affect us and how we perceive them, they still offer little insight into how we develop relationships with technology. Indeed, Hassenzahl's framework can, for instance, reveal that voice lighting control is pleasurable, and Mekler and Hornb{\ae}k's can be used to show how accessing audiobooks with Alexa can provide a meaningful experience. Yet, we still do not know how to describe the qualities of an experience that can lead to Alexa assuming a social role~\cite{Purington:2017:AMN}. 

To find explanations for phenomena such as the aforementioned one, we turn to literature on psychological needs. Psychological needs have previously been used as a means of explaining the user experience of interactive technologies~\cite{Hassenzahl:2010:NAI:1837536.1837814}. 
\subsection{Psychological Needs and Interactive Artefacts}
Almost two decades ago, Jordan~\cite{jordan2002designing} introduced his framework of the four pleasures, physio-, socio-, psycho- and ideo-pleasure to the research community. Jordan's work is inspired by Maslow's hierarchy of needs~\cite{maslow1943theory}. Jordan emphasised the need to integrate pleasure-based approaches into interactive products, and, most importantly for this work, stated that his framework introduced pleasures people might seek as well as pleasures products can potentially address. Physio-pleasure is determined by the sensory organs and connected to physical sensations. Socio-pleasure can be described as the joy or the satisfaction that arises through social interactions. Psycho-pleasures are pleasures connected to cognitive as well as emotional reactions. Pleasures stemming from people's values are called ideo-pleasures.

A year later, Sheldon et al.~\cite{sheldon2001satisfying} published their work on psychological needs  (see table~\ref{tab:needs} for an overview of the different psychological needs). Their study is of particular importance for the HCI community since one of the most influential approaches to psychological needs in interaction design is based on their work. More precisely, Hassenzahl et al.~\cite{Hassenzahl:2010:NAI:1837536.1837814} emphasised the importance of psychological needs for experience-oriented technology design. Sheldon et al.~\cite{sheldon2001satisfying} conducted three consecutive studies in order to explore the importance of a variety of psychological needs. They derived and explored ten psychological needs from a variety of theories of psychological need fulfilment: self-esteem, autonomy, competence, relatedness, pleasure-stimulation, physical thriving, self-actualization-meaning, security, popularity-influence and money-luxury (see table \ref{tab:needs} for an overview and short descriptions). The four most salient needs in their study were self-esteem, autonomy, competence and relatedness, which is in line with and extends Self-Determination Theory by Ryan and Deci~\cite{ryan2000self}.

In psychological research with a focus on heodnia and eudaimonia, meaning is one of the four core definitional elements (i.e. authenticity, meaning, excellence, growth) that encompass eudaimonia~\cite{huta2014eudaimonia}. Furthermore, psychological research found a bidirectional connection between psychological need fulfilment and the concepts hedonia and eudaimonia, e.g.,~\cite{saunders2018physical}. This is in contrast with Mekler and Hornb{\ae}k~\cite{Mekler:2019:FEM:3290605.3300455}'s work where the concept of eudaimonia was not used in the framing of meaning. This suggests that there is a need to further explore the concept.

\begin{figure}
    \centering
    \includegraphics[width=.95\columnwidth]{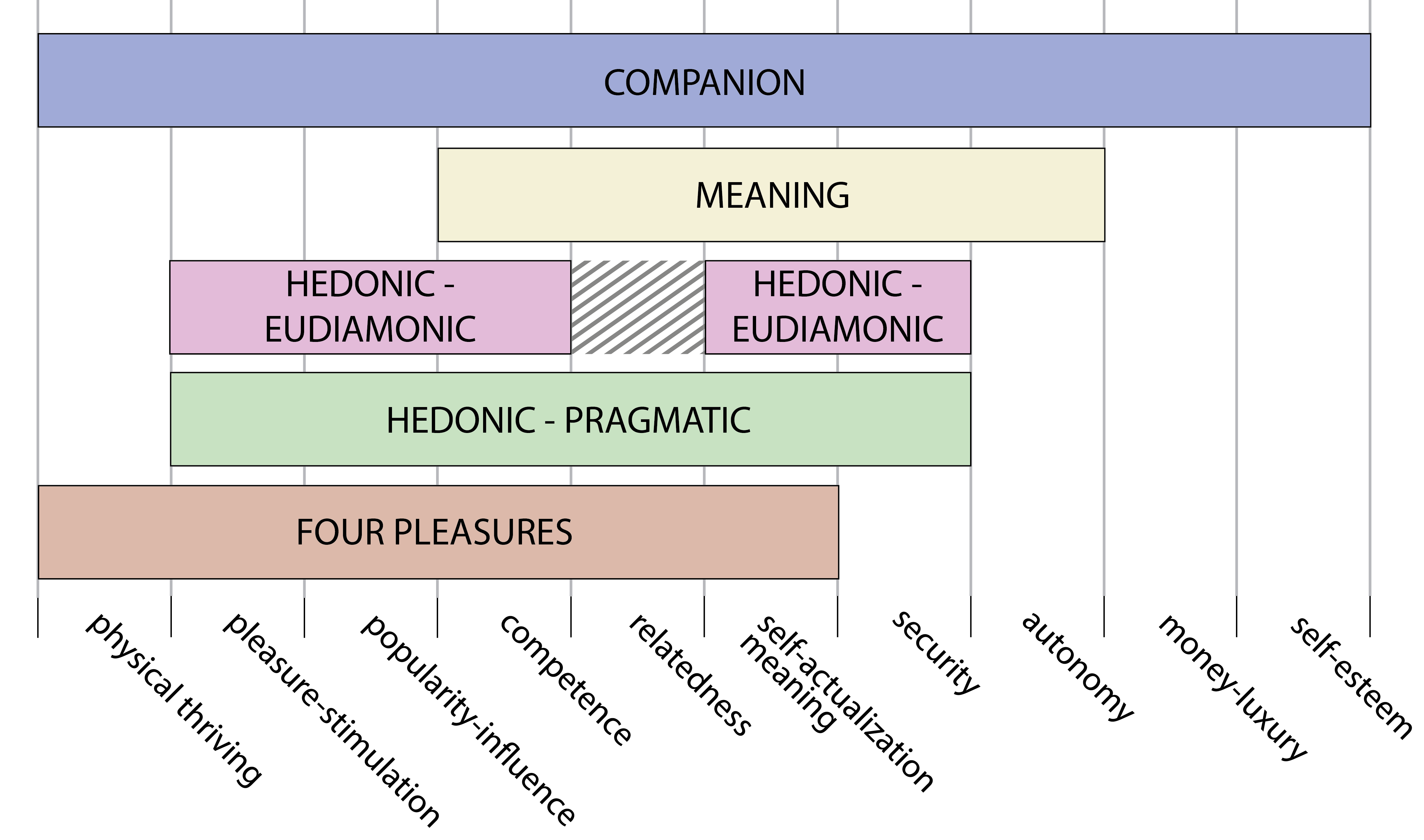}
    \caption{Psychological needs and the corresponding experiences. Mapping of the four pleasures and hedonic-pragmatic based on~\protect\cite{Hassenzahl:2010:NAI:1837536.1837814}. Mapping of hedonic-eudaimonic based on~\protect\cite{Mekler:2016:MPL:2858036.2858225}. Note that the mapping of meaning is based on the interpretation of the authors of this work, as no explicit relation to the full spectrum of needs was provided in the original paper~\protect\cite{Mekler:2019:FEM:3290605.3300455}.}
    \label{fig:grid}
\end{figure}

The avid reader can observe that most of the psychological needs proposed by Jordan or Sheldon have been addressed by past conceptualisations of user experience (see figure \ref{fig:grid}). Yet, none of the current user experience frameworks directly addresses needs related to the social aspects of interactions with a technological device as opposed to the social aspects of interactions with humans mediated through a technological device. Traditionally, the social aspect of interactive technologies has been limited to multiple users interacting with or through an artefact. The domain of social computing specifically studies how social behaviour is affected by computer technology. 

However, what happens when technologies become social actors? While we have known since early HCI days that technologies can have social features~\cite{Nass:1993:AAE}, how do we understand the way in which they weave themselves into the fabric of our social being? To begin answering this question, we propose conceptualising artefacts that have a profound social presence as \emph{companion technologies}. Before we present our notion of companion technologies, we need to take stock of the current sate of the art in HCI on understanding socially embedded technologies.

\begin{table}
    \centering
    \begin{tabularx}{\columnwidth}{>{\hsize=.4\hsize}XX}
    \toprule
    Psychological need         & Description \\ 
    \midrule
Autonomy                   & Feeling that activities are chosen by oneself and to be the agent of one's life  \\ 
Competence                 & Feeling effective and competent                                                      \\ 
Relatedness                & Feeling close to some other individual, feelings of interpersonal connection        \\ 

Self-actualization-meaning & Sense of long-term growth                                                            \\ 
Physical thriving          & Sense of physical well-being                                                         \\ 
Pleasure-stimulation       & Feeling of pleasurable stimulation                                                   \\ 
Money-luxury               & Focus on wealth, luxury and nice possessions                                          \\ 
Security                   & Feeling of order and predictability in one's life                                    \\ 
Self-esteem                & Self-worth and global evaluation of oneself                                          \\ 
Popularity-influence       & Feeling of having the ability to win friends and influence people  \\
\bottomrule
    \end{tabularx}
    \caption{Psychological needs and brief descriptions based on the work from Sheldon et al.~\protect\cite{sheldon2001satisfying}. Note that their work integrated a variety of different needs from other works, such as Deci and Ryan's self-determination theory~\protect\cite{ryan2000self}, Maslow's theory of personality~\protect\cite{maslow1943theory} and Epstein's cognitive-experiential self-theory~\protect\cite{epstein1998cognitive}.}
    \label{tab:needs}
\end{table}{}

\section{What Are We?: Concepts of Social Agent Technologies in HCI}
In HCI, socially embedded technologies have been explored under a variety of different names. Norman~\cite{Norman:1994:MPI:176789.176796} discussed the chances, challenges and myths connected to social agents. Shneiderman and Maes~\cite{Shneiderman:1997:DMV:267505.267514} discussed the idea that humans will have to entrust specific tasks to digital agents. These technologies should then either act on a user's behalf or offer suggestions to the user. Today, the scenario described by Shneiderman and Maes is reality. Artificial agents have become ubiquitous, but the interaction between humans and computers is still in flux. Farooq and Grudin~\cite{Farooq:2016:HI:3012754.3001896} addressed this shift by discussing the continuum from Human-Computer Interaction to Human-Computer Integration. They stated that the interaction between humans and computers could be described as stimulus-response and emphasised that this interaction style does not constitute a human-computer partnership. On a similar note, the notion of computers as equal partners has been a longstanding subject of critical debate within the HCI community~\cite{Farooq:2017:HCI:3027063.3051137}: What constitutes a meaningful human-computer partnership that takes user needs into account? Should computers try to understand users in real time? Is an in-depth understanding of internal processes of the machine needed? Do humans always keep control of the actions of the technology? All these questions continue to be discussed, and the notion of computers as partners is more relevant than ever~\cite{Rozendaal:2019:OID:3341168.3325277}.   

HCI's reaction to a proliferation of social agents around us has been a very analytical one; an attempt at classifying such artefacts. Digital game research analysed the design space of companion characters in games~\cite{10.1145/3242671.3242709}. Grudin and Jacques~\cite{Grudin:2019:CHQ:3290605.3300439} outlined the design space for conversational agents and discussed the term \emph{human-computer symbiosis}. Their work focused on chatbots that engage in conversations and categorised them. Interestingly, the paper points to contradicting empirical results regarding successful chatbot design. Based on the example of embedding humour in the design of a chatbot, the authors emphasised the difficulty when it comes to designing `good' bots. While some users enjoyed the humorous responses, the results also showed that humour led to higher expectations towards the technology. In other words, the designer is faced with the conundrum of personality characteristics increasing the potential of positive experiences for users when interacting with chatbots, which in turn may lead to expectations that cannot be met. This issue illustrates how we need to better conceptualise technologies which are social agents in order to design them successfully.

Past research has also addressed specific types of agents. `Objects with Intent' was one proposed concept which included intelligent everyday things such as lamps or jackets~\cite{Rozendaal:2019:OID:3341168.3325277,Rozendaal:2016:OIN:2930854.2911330}. In his recent study, Rozendaal~\cite{Rozendaal:2019:OID:3341168.3325277} found that the same objects can be framed as a tool and as a partner. He listed qualities agents usually possess, such as being social, acting autonomously, being reactive, and acting proactively. Further, unlike in relationships and collaborations between two human partners, humans have the ultimate control in human-object partnerships. Rozendaal's classification offers an understanding of agents on a functional level and a discussion of the perception of artefacts on an experiential level. What Rozendaal does not address is a holistic understanding such artefacts on an emotional and experiential level, one that goes beyond a specific use case or study. For instance, we argue that humans might not always want to be in control; that they actually enjoy passing on some of the responsibility to an intelligent, trustworthy agent.

Given that artefacts with social agency have already been widely discussed and studied, why do we need yet another term? First, while an analytical approach is often worthwhile, a meta understanding of the problem is also required. Contrary to past efforts that attempted to classify artefacts with social presence, we suggest focusing on the experiential qualities of such artefacts per se. Further, we agree with Grudin and Jacques~\cite{Grudin:2019:CHQ:3290605.3300439} that managing expectations is key in interacting with agents. Thus, we need to understand our experience of socially embedded technologies to know what we expect of them. 

Our past experiences shape who we are~\cite{fuster2015past}. Consequently. if a user projects a personality onto an interactive artefact, this projection stems from the user's own experiences. As expectations of such a projection are inherently tied to past experiences, projecting social agency onto a technology must lead to empathy for the object. In other words, if a person projects something onto an artefact (e.g. a personality), this projection is intertwined with their previous experiences. Thus, empathising with the projected personality is not a possibility but a given, since the projected personality emerged from the experiences of the person who projected it onto the artefact in the first place. This observation is the key notion of this theoretical exploration.  However, expectations are not only set by previous experiences that are directly related to the current situation. For instance, users who interact with an innovative, novel technology for the first time do not have any experience directly related to the interaction with this novel technology. Instead, their expectations can stem from previous experiences that go beyond directly related incidents. For instance, the aesthetic appearance of the technology could remind them about something they experienced in the past. The (re)interpretation of this past experience combined with the perception of the current situation could then potentially lead to the person projection something onto the technology that is, as mentioned above, tied to past experiences~\cite{fuster2015past}.

\section{Companion Technologies}
Our analysis shows that while past work addressed many notions that help us understand socially embedded technologies, there is a need for a meta-approach that would address, inter alia, the full scope of psychological needs, the projection of personality onto objects, the emergence of the computer as a partner and the intentionality of objects. Consequently, we turn to the meta-science of philosophy to propose the concept of \emph{companion technologies}, which is embedded in the philosophical notion of empathy for objects.


In order to unpack the notion of empathy for HCI, we propose to use the term \emph{companion technologies}. A companion technology is an interactive artefact that can evoke empathy in a user.

The philosopher and poet Johann Gottfried Herder discussed differences in the experience and the perception of different objects such as sculptures and paintings~\cite{herder2002sculpture}. He argued that sculptures are made for the tactile experience, whereas paintings are made to be looked at. Consequently, we need to ask what the purpose of a companion artefact is. Are they made to be looked at, touched, or experienced? We propose framing companion technologies as artefacts that are designed to evoke an emotional response, similar to that of interacting with another human.

We postulate using the word `companion' to stress the 
relationship between the artefact and the user. The etymology of the word (`one who breaks bread with another') stresses a deep mutual relationship and a complementary duality. It emphasises the embedding of the artefact in the everyday life of the user. This contrasts with the term `partner', which etymologically emphasises division and evokes connotations with work environments. The term `agent', stresses acting on the world and empathy does not necessitate active participation.

\section{Empathy for Objects in Philosophy}
To fully understand companion technologies as ones that evoke empathy, we need to first analyse the concept of empathy. As part of our theoretical exploration, we review work in philosophy about empathy which people can develop towards artefacts. The origin of the word empathy is the Greek empatheia (`em' means `in' and `pathos' means `feeling'). Empathy is `the ability to share someone else's feelings or experiences by imagining what it would be like to be in that person's situation'. Early work by Herder~\cite{herder1778erkennen} described what was later dubbed empathy as the ability of individuals to understand, feel nature with all its manifestations in analogy to oneself and perceive these manifestations.

Through empathy, individuals make sense of other minds~\cite{stueber2010rediscovering}. In the beginning of the 20th century, empathy was a combination of two independent philosophical traditions; philological sciences and philosophical aesthetics. While the philological sciences mainly focused on the notion of understanding (`Verstehen'), philosophical aesthetics introduced the concept of empathy (Einf\"{u}hlung) per se~\cite{stueber2010rediscovering}. 

Theodor Lipps~\cite{lipps1906einfuhlung} united these two notions. For him, `Verstehen' and `Einf\"{u}hlung' are connected as they both have to do with how we understand phenomena that express themselves externally, but at the same time represent an internal expression through their external appearance. For instance, a mental state can be expressed through an artefact (Einf\"{u}hlung) or through a physical reaction (Verstehen). Due to this affinity, these two notions have sometimes been used interchangeably. Lipps argued further that empathy is based on the mysterious tendency of humans to motor mimicry. He postulated that, since this tendency is often not allowed, for instance due to external circumstances or social norms, this interdiction leads to `inhibited imitation'; an inner tendency to imitate. This led him to explore the things and creatures which we can imitate and ask if something or someone is a minded creature or a minded object~\cite{stueber2010rediscovering}. This dilemma is still relevant today as we wonder if and when we can ascribe social agency to objects.

Lipps was criticised by Edith Stein~\cite{stein1917problem} for using the notions of empathy (`Einf\"{u}hlung') and feeling of being one with the other individual or an object (`Einsf\"{u}lung') interchangeably~\cite{stueber2010rediscovering}. He reacted to this criticism by explaining that empathetic identification did not lead to the person losing themselves. Instead, it could be compared to feeling the sadness communicated through a piece of art; one can feel the sadness, but without all its motivating force~\cite{lipps1906einfuhlung}.

The thinking of Lipps was critically questioned and further developed by Stein and Edmund Husserl. They extended the understanding of empathy from understanding other minds to supporting personality development due to the ability to acknowledge, engage and understand opinions other people have about oneself~\cite{hauser2003lotze}. This idea echoes our experiences with interactive agents. Individuals can feel empathy towards companion technologies, but the companion technology can potentially also trigger personality development through taking an independent position and confronting the user with content (or potentially opinions) about him or herself. 

Rudolph Hermann Lotze, one of the most important philosophers of the nineteenth century, had a significant impact on Lipps's work. Lotze and Lipps had a common view of a concept core to HCI---experience. They defined two different kinds of experiences: sense-experience and feeling-experience~\cite{frechette2013searching}. Sense-experience is object-directed, whereas feeling-experience is self-directed. In the description given by Lipps, experiences always embed two perspectives an immediate perspective and a mediate perspective~(as cited in~\cite{frechette2013searching}).

Even though sometimes critically discussed, the aftereffects of Lotze's thinking can be found in neo-kantianism, phenomenology, Frege's conception of logic, psychology and theology~\cite{hauser2003lotze,frechette2013searching}. Lotze sees knowledge not as one system but as a variety of different views. This view resembles the eclectic and multidisciplinary nature of knowledge in HCI. In his work, Lotze defines beingness (`Das Sein') as relations; relations as a system of interactions. Through the expression of such relations in the soul, meaning and significance become apparent. In other words, Lotze is arguing for the predominance of the subjective mind, where all phenomena take place. The subjective mind must strive to be objective and its integration into reality is imperative. As we argue further below, this stance is also applicable to companion technologies in HCI. 

Lotze addressed and attempted to resolve the conflict between scientific realism and idealism. He emphasised the enriching the reciprocal relationship between abstract constructs of idealism and applying them to the world. On the other hand, he remerked that the adoption of principles focusing on the behavior of things (taking a scientific realism stance) and assuming their validity requires idealism to some extent. We call this complementary duality. Complementary duality can be a useful tool to shed light on the duality regarding the understanding of companion artefacts in HCI research. Adopting a complementary duality mindset disentangles the differentiation between empathy for objects and the inevitable experiential companion qualities that correlate with it.

Another key differentiation in Lotze's work is between `Vorstellen'---envisioning as a cognitive function and `Vorgestelltem'---the object which is envisioned. This differentiation points towards the distinction between how a digital companion is experienced and what it actually is. While the user and the companion technology are in a mutual relationship, the user is the one in the active role, projecting social features onto the artefact.

Lotze argues further that, when it comes to deciding between different alternatives, it is not enough to consider which one of the different alternatives is necessary. Instead, one also should consider what is meaningful. Applying this thinking to companion artefacts facilitates an understanding of how they can be designed and which experiential qualities might be embodied in them. Here, Lotze's views echo the framework proposed by Meckler and Hornb{\ae}k~\cite{Mekler:2016:MPL:2858036.2858225}. Furthermore, Lotze stressed that this differentiation was not synonymous of a value-system but instead was value-neutral. This is congruent with HCI's pursuit to build utilitarian artefacts that also provide meaning and pleasure.

\section{An Experiential Framework for Companion Technologies}
\begin{figure}
    \centering
    \includegraphics[width=.95\columnwidth]{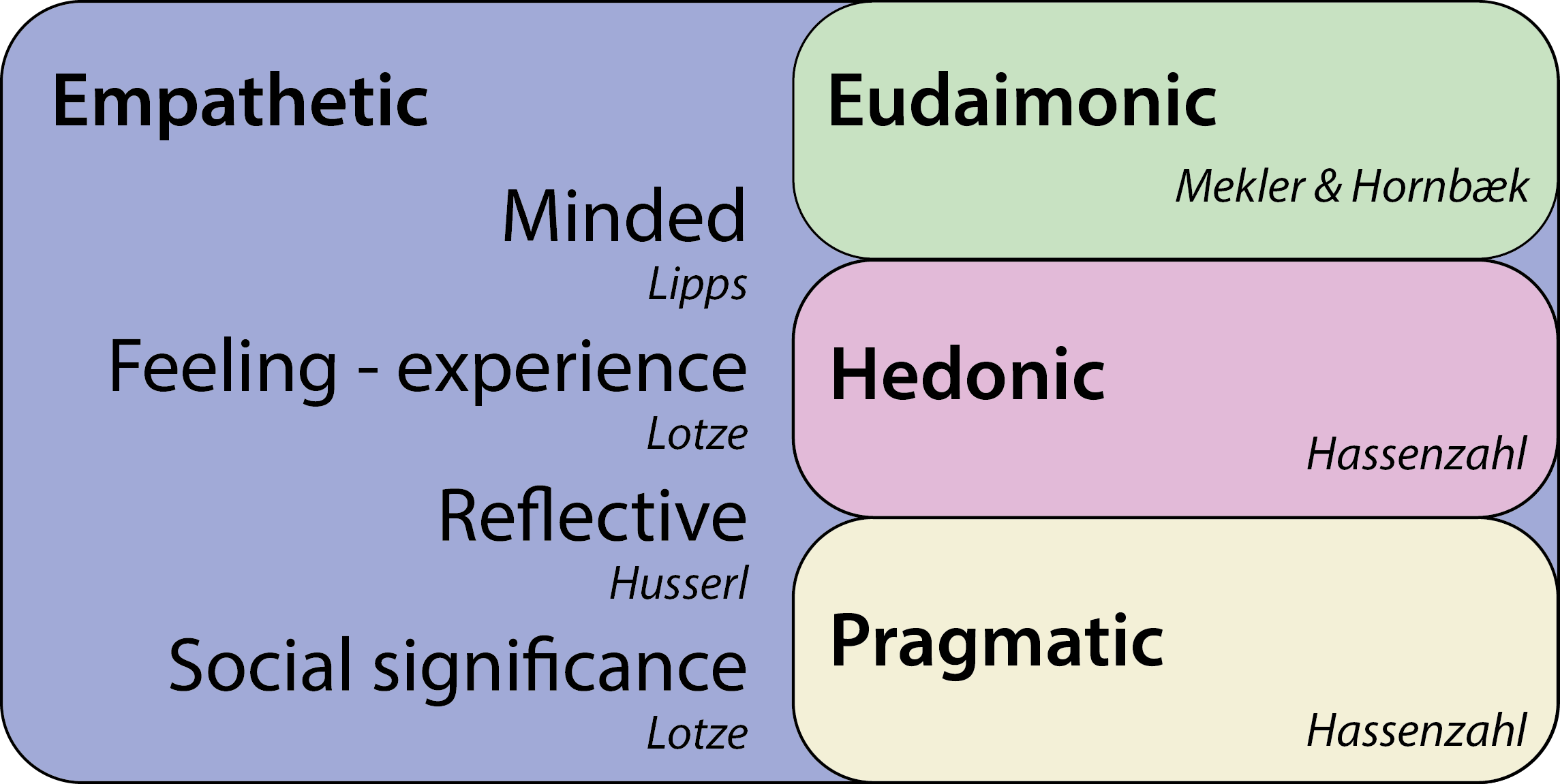}
    \caption{The Experiential Framework for Companion Technologies. The framework extends our understanding of experience of technology by introducing empathetic experience. We use four concepts from philosophy of empathy to define empathetic experience.}
    \label{fig:framework}
\end{figure}

Having demonstrated how work in the philosophy of empathy aligns with the pursuits of HCI, we propose a way of conceptualising companion technologies and understanding the experiences which companion technologies evoke. Importantly, our goal is to capture the felt experience of companion technologies and not their true nature. We are not focusing on functionalities or technological characteristics of artefacts and how they might be manipulated to create a desired reaction in users' (e.g. empathy). 
Our framework explicitly extends past understandings proposed by Hassenzahl~\cite{hassenzahl2003thing} and Mekler and Hornb{\ae}k~\cite{Mekler:2016:MPL:2858036.2858225} by including the social aspects of the experience between human and companion technology. Such social aspects are crucial for companion technologies. We specifically focus on social experiences that do not occur between two humans mediated by technology. Instead, we study social experiences between the human and companion technology. Figure~\ref{fig:framework} illustrates the framework.

Before we describe the framework in detail, we must mention its key limitation. As Stueber~\cite{stueber2010rediscovering} noted, Lipps's hypotheses regarding empathy are neither scientifically sufficiently explored nor empirically proven. In line with the critique outlined by Stueber, we do not argue that the notion of empathy based on Lipps's understanding can necessarily lead to genuine insights about the mental state of other individuals. On the contrary, we emphasise that it is unimportant (from an experiential standpoint) if individuals generate what they might deem true or objective knowledge about other individuals or objects. Instead, we extend the current understanding of subjective user experience and what makes interactions with technologies meaningful. We attempt to integrate empathy towards objects into the current understanding in HCI. Experiential qualities caused by empathy towards objects do not stem from real insights or true knowledge, but are sourced from the innate human need to feel someone or something.

\subsection{The Four Concepts}
In line with the philosophy of empathy, we call the dimension of experience specific to companion technologies \emph{empathetic}. Empathetic qualities of technologies are complementary to pragmatic, hedonic and eudaimonic qualities. Further, we synthesise four concepts from philosophical work that help in identifying empathetic qualities in technologies. In the following, we describe these concepts in detail. For each of the concepts, we list challenges for HCI that emerge from the conceptual stance taken in our framework.

\subsubsection{Minded}
Firstly, we postulate that companion technologies which evoke empathy must be minded. This concept is inspired by Lipps's investigations (cited by~\cite{stueber2010rediscovering}) into what constitutes minded creatures or minded objects. Building on Lipp's thinking, humans can experience technology empathetically if a companion technology produces a sense of wonder. Such a sense of wonder could stimulate people to reflect about questions such as: Is this creature (i.e. technology) minded? What is the emotional state of this artefact? 
If an artefact is perceived as supernatural or mystical, users try to make sense of it, and strive to understand if it is minded. That is where the empathetic experience begins. Here, wondering about the technology goes beyond wondering about how the technology works, beyond the technical aspects of it. The technology produces a sense of wonder in users which evokes empathetic experiences because of the need of users to understand its mind. The spectrum of wonderment about companion technologies ranges from joy because the companion is perceived as amusing or helpful (e.g. mobile digital assistants), through surprise about the efficiency of the technology (e.g. smart home appliances), to awe because of the incomprehensible sophistication and cognitive abilities of an android (e.g. encountering a life-like social robot). The fascination with the technology paired with a magical element in the interaction process leads users to ask questions about themselves and the interactive artefact. Only a companion that is perceived as minded by the user can form a social dyad with the user. This, in turn, leads to the user wondering about the motives of the companion and how the companion perceives the world, thus forming an empathetic bond with the technology. Consequently, the following challenges for HCI arise:

\textsc{Future challenges for HCI:} How do we design for mindedness of companion technologies? How can we inform users about the mindedness of specific companion technologies? Which design characteristics lead to different perceptions of minds of technologies as companions?

\subsubsection{Feeling-Experience}
A true companion technology enables the user to be part of the technology's feeling-experience (Selbstgef\"{u}hl, as postulated by Lotze~\cite{frechette2013searching}). According to Lipps, sense-experiences are object-directed, while feeling-experiences are self-directed. For instance the feeling-experience of bodily warmth is related to the sense-experience of the heat of the radiator. Yet, feeling-experiences can also be object-directed, e.g. the heat of one's body. On the one hand, this implies that users should be able to establish boundaries between their perceptions of themselves and companion technologies. On the other, users are likely to project the ability to have both kinds of feelings onto companion technologies and expect the same of the technologies. This is perhaps best illustrated using an example. The smart oven that will switch off when the user is manipulating food inside of it builds a perception of knowing the user's body heat. As the oven also communicates its own heat, the user may perceive the experience as empathetic because the oven and the user share a feeling-experience. Hence, the quality of the interactive artefact being able to present itself as capable of having feelings that are not object-directed builds an impression of agency for companion technologies. As outlined above, we assume that people potentially enjoy to delegate some responsibility to a companion technology with agency. The question remains: how can this need be translated to the design of technologies as companions?

\textsc{Future challenges for HCI:} How are different levels of agency and different levels of fidelity of companion technologies related? How can we balance companion agency and the companion assuming responsibility for tasks?

\subsubsection{Reflective}
Having established that a true companion technology is perceived as minded and capable of self-directed feelings, we now turn to the feelings that the technology evokes in users that build empathetic experience. The need to wonder about technologies and investigate if they are minded is motivated by the fact that minded entities are able to express opinions and thus may have opinions about the user. Here, we apply Husserl's thinking and stress that the empathetic experience is built by the companion technology's perceived ability to form and engage with opinions about the user. Consequently, how a person perceives their companion might also change how one perceives themselves, thus giving the companion social agency. Through these interactions and because of developing an empathetic relationship, users have the potential to learn and develop based on the interaction with the technology. This is where our framework connects with eudaimonic experience, creating a continuum of experiences.

\textsc{Future challenges for HCI:} How do we design for reflective experiences with companions that motivate users to learn about themselves and develop their potential? How can we prevent creating digital companions from exhibiting behaviours which users perceive as judgemental?

\subsubsection{Social Significance}
Finally, we use Lotze's notion of significance to highlight how companion technologies can assume social roles. Through the expression of relations in the soul, meaning and significance become apparent. In order for an artefact to evoke empathy, the relation to the object must be perceived as significant. As a user integrates the technology into his or her everyday life, they intentionally ascribe meaning to the actions of the objects and thus give significance to the object's actions. Stueber~\cite{stueber2010rediscovering} implied that empathy is the default method to build an understanding about other minds. We extend this position and argue that empathy is central to building an understanding of inanimate objects as well as of the bonds people develop towards them. Thus, empathy towards objects can be the means to explain the inexplicable. This fact is of increasing importance as systems become more and more complex.

In this concept, our framework is also inspired by Rozendaal's work~\cite{Rozendaal:2019:OID:3341168.3325277} who outlined and applied Dennett's theory of intentionality. The notions of the theory of intentionality complement the philosophy of empathy, not least because of the conceptual affinity to folk psychology. 
Dennett's theory of intentionality includes three stances: the physical stance, the design stance and the intentional stance. This can be explained using the example of a sundial, where one could say that a sundial tells the time because it combines the Sun's altitude or azimuth with the gnomon and makes a shadow. Seen from a design stance the sundial shows the time with the help of the sun because it was designed to do so. From an intentional stance, the sundial has beliefs about time, and acts on its beliefs because of a desire to show a specific time dependent on the position of the sun. Similarly, the theory of intentionality explains the psychology of how users can ascribe social significance to companion technologies.

\textsc{Future challenges for HCI:} How can we manage the user's understanding of the companion technology and enable an evolution of understanding? How do we assure that object stay mystical enough to remain significant? How do we design technologies that take actions to which users can ascribe meaning?

\section{Where Are We Going?: Using the framework}
So far, our framework may seem ephemeral as it is derived from high-level concepts. To mitigate this, this section discusses two examples from previous research on technologies that may be companions. We show how our framework enables analysing the experiences reported in the research and mapping the empathethic experiences of the technologies in question.

We chose two examples of past research~\cite{Sung:2007:MRR:1771592.1771601,Cho:2019:OKF:3322276.3322332} that investigated artefacts that can be conceptualised as companion technologies. These two papers address two companion technologies that are present in many modern households. We chose a robotic vacuum cleaner and a smart speaker as the two examples. These companion technologies have different levels of complexity, offer different functionalities and use different interaction techniques. Yet, they share many companion qualities. While we cannot pay justice to the two papers and analyse them in full depth, we provide an overview of how the results presented in those works could be interpreted with our framework. We highlight the psychological needs that can be identified in the two examples through the application of our framework in bold font. Please note that we highlight psychological needs in case they were fulfilled or the potential for their fulfilment could be identified.

\subsection{Roomba}
We first take a retrospective look at work by Sung et al.~\cite{Sung:2007:MRR:1771592.1771601} who investigated how users developed relationships with their robot vacuum cleaners and the impact the device had on the home. A key finding of the work was that users developed intimacy towards their cleaning robots (\textbf{Relatedness}). Through the lens of our framework, intimacy towards objects can be understood as part of the empathetic experience. The study collected and analysed postings from roombareview.com and conducted follow-up interviews with 30 participants. Sung et al. described their participants as Roomba enthusiasts. Based on their data analysis they derived three themes, \textit{Feeling happiness towards Roomba}, \textit{Lifelike associations and engagements with Roombas} and \textit{Valuing Roomba: Promoting and protecting it}. Their findings can be explained through the lens of companion technologies.

In general, Sung et al.~\cite{Sung:2007:MRR:1771592.1771601} found that participants formed stable, intimate attachments to their Roombas (\textbf{Relatedness}). This finding can be explained through our framework concept of \textsc{social significance}. Through integrating the Roomba into their everyday life, it became a significant part of it. Within the theme \textit{Feeling happiness towards Roomba}, participants described situations where the Roomba forced the whole family to be neater. This example showcases the learning experience (\textbf{Self-actualization-meaning}) we describe in the framework concept \textsc{reflective}. The owners of the Roombas gave the object social agency, thus valuing the opinion of the vacuum companion. In a sense, the Roomba formed the opinion about its owners that they were untidy. Users engaged with the companion technology and valued the opinion of the artefact. Consequently, the whole family learned and became neater.

Interestingly, the study by Sung et al. also showed that the participants were positive towards the reflection and learning process triggered by the devices (\textbf{Self-actualization-meaning}). In contrast, the positive experiences with their Roomba provided a balance for extra work required for behaviour change. This shows the differentiation we addressed in \textsc{feeling-experience}. The study participants had two self-directed experiences; one was the experience of not enjoying tidying that much, the other the experience of enjoying the presence of their Roomba a lot (\textbf{Pleasure-Stimlation}). At the same time, their Roomba had agency, making them neat and content. Thus, the user and the Roomba shared a feeling-experience of tidying and sharing joy (\textbf{Relatedness}).

Further, the results suggested the \textsc{social significance} of the Roomba. The theme \textit{Lifelike associations and engagement with Roomba} showed that the participants formed an understanding about the object and ascribed intentionality to it. For instance, participants gave their vacuum robots names, nicknames or even changed the name after a while so that the name fit the personality of the Roomba. Furthermore, some described it as a valuable family member and one participant stated that he felt a stronger bond to his Roomba than to his mopping robot (\textbf{Relatedness}). This showcases that through the integration of the object into the users' everyday lives, the vacuum robot assumes social roles and its \textsc{social significance} increases.  

As the Roomba changed from an inanimate artefact to `something like a pet' or a `family member', users developed a sense of pride for the device. The results from Sung et al. showed that participants were proud to own one (\textbf{Money-luxury}). Similar to pet owners often showing pictures of their cat, Roomba owners showed the robot to friends and family, wrote emails about its positive characteristics and worried about it when it had functionality issues (`the Roomba is sick', thus capable of a \textsc{feeling-experience}).

To summarise, the findings by Sung et al. can be described through the lens of our framework. The empathy which participants have shown towards their Roomba is reflected in the described experiences as well as in the choice of words of the participants in the study.

\subsection{Amazon Echo}
Next, we use our framework to understand the results of a more recent paper. Cho et al.~\cite{Cho:2019:OKF:3322276.3322332} explored how eight households used an Amazon Echo over 12 weeks. The authors described the journey of the study participants and their Amazon Echo. They mapped their results onto five stages, which represent the experiential journey of owning the device: \textit{pre-adoption}, \textit{adoption}, \textit{adaptation}, \textit{stagnation}, \textit{acceptance}.  In contrast to Roombas above, this research shows the Echo did not become a companion technology for the study participants (\textbf{Relatedness}) as the devices were eventually abandoned. Nevertheless, our framework can be applied and generate insights, independent of the success of an interactive artefact becoming a companion technology or not. We illustrate this in the following paragraphs.

Initial results of the study showed that Alexa may have gained \textsc{social significance}. However, the users soon started to feel disappointed about the device. The technology did not live up to their expectations. In the view of our framework, this implies that users had the expectation of Alexa being \textsc{minded}. In the beginning of the study, some participants described a sense of joy when they interacted with Alexa (\textbf{Pleasure-stimulation}), as described in our framework concept \textsc{social significance}. But the sense of joy did not evolve to wonderment about the technology nor did it offer reflective experiences (\textbf{Self-actualization-meaning}). The participants did not enjoy interacting with Alexa (\textbf{Pleasure-stimulation}) and the usefulness score for the Amazon Echo was lowest in the adoption phase.

At first glance, Alexa appears to be more sophisticated than the vacuum robots from the previous example. But, contrasting the two studies, users developed a more empathetic bond towards their Roombas compared to Alexas. One potential explanation behind this difference, based on our framework, could be that the Amazon Echo is static, whereas the Roomba conquered personal space in the houses and apartments of the participants. Consequently, people started to adjust their routines and their space to the needs of the technology and tried to make sense of the Roomba's random movements~\cite{Sung:2007:MRR:1771592.1771601}. In contrast, participants were not `forced' to make sense of Alexa. They simply put the device in a corner and were able to forget about it. This is reflected in the statement of one participant in Sung et al. study who pointed out that he `simply forgot about her'. As most participants did not value Alexa, it failed to become \textsc{socially significant}.

One possible explanation behind Alexa's failure could be that, in line with previous work and results from Cho et al.~\cite{Cho:2019:OKF:3322276.3322332}, voice interaction is connected with users perceiving technology as human-like. This raises their expectations towards the technology. We hypothesise that due to this fact the expectations towards Alexa providing a companion-like experience are higher than towards Roomba. Users had to make a larger effort to make sense of Roomba because the robot did not use human speech but beeping sounds instead~\cite{Sung:2007:MRR:1771592.1771601}. The need for a \textsc{minded} Alexa was further strengthened by the reported desire for a smart home. In the view of our framework, this could mean that participants would have enjoyed to delegate some responsibility to the digital companion (\textbf{Autonomy}), which is addressed by the \textsc{Reflective} concept in our framework. The contrast between these two cases shows that the forming of an empathetic bond between the user and their (prospective) companion technology is key for sustained use that brings joy. Failure to form such a bond resulted in abandonment.

\section{Companion Technologies as a Complementary Concept}
While we do believe that our philosophy-driven approach to companion technologies offers new ways of interpreting how users interact with everyday objects, we recognise that our framework is a high-level concept. Here, we discuss how companion technologies complement other concepts, allowing for a broader scope of inquiry. We suggest that future studies of technologies that may be companions use our framework together with more specific concepts to gain a thorough understanding of the overall experience.

There is conceptual overlap between individual components of our work and research by Mekler and Hornb{\ae}k~\cite{Mekler:2019:FEM:3290605.3300455}. However, the main differentiation between their work and ours is on the meta-level. Mekler and Hornb{\ae}k~\cite{Mekler:2019:FEM:3290605.3300455} focus on the experience of meaning in HCI (e.g. feeling connected to oneself and the world mediated by technology). In contrast, our framework describes the social bonds between humans and their companion technologies. Hence, the interactive artefact evolves from a technology that mediates connectedness between humans to a companion technology humans can relate to and feel connected with, without a third party involved. 

Symbiotic interactions~\cite{jacucci2015symbiotic} take a sensing-centered approach to designing companion technologies. The intended role of our work is providing an alternative perspective on the experience of technologies that evoke empathy by adapting the philosophical work on empathy for objects, thus allowing HCI researchers and practitioners to use philosophical knowledge to address HCI issues. Thus, the focus of our framework is on a persons' experience of a technology, rather than characteristics internal to it. Symbiotic technologies can be combined with our approach to study the latter aspect.

A body of work adopted a psychological approach to the issues addressed in this paper. There is a significant body of empathy research in social psychology (e.g.~\cite{davis2018empathy}) that focuses on empathy in human relationships. Also, early HCI research addressed the notion of computers exhibiting features perceived as having personality~\cite{nass}. Based on our framework, technology can be perceived as having a personality. However, we focus on empathetic bonds between humans and technology with a wider scope, where perceived personality can be one of a variety of contributing factors. Later work by Turkle~\cite{turkle2004whither} showed that users exhibited the ability to see computers as \emph{second selves} or even attribute \emph{states of mind}~\cite{turkle} to computer systems. Turkle et al.'s work provides a psychological understanding of the underlying mechanisms of how user socially relate to computer artefacts. They empirically determined that artefacts can \emph{elicit admiration, loving and curiosity}. Our work complements Turkle et al's understanding by addressing the phenomenological concerns behind companion technologies and offering an explanation beyond psychology. Our work explains the concepts behind what Turkle at al.~\cite{turkle} called a \emph{state of mind} through empathy and the related concept of self-mindedness.

Furthermore, our findings are related to certain concepts used in the field of Human-Robot Interaction (HRI). Alves-Olivieira et al.~\cite{AlvesOlivieira2019} experimentally manipulated empathetic actions (e.g. a robot recalling previous experiences or reacting to emotional states of participants). In contrast, our work does not focus on empathetic interactions evoked by specific functionalities of a certain technology. Instead, this work focuses on empathetic bonds formed due to intrinsic, reflective processes, independent of specific functionalities of the interactive technology.  
Furthermore, we acknowledge the conceptual closeness of the notion of \textit{Empathy for Objects} to Theory of Mind~\cite{premack1978does}, which was used to interpret user interactions with social robots, e.g.~\cite{Levin:2009:DDS}. Whereas we focus on emotional experiences embedded in the notion of empathy, theory of mind focuses on the cognitive perspective. In a initial philosophical investigation, Kahn et al.~\cite{kahn} discussed the possible emergence of new ontological categories represented by social robots. The inquiry into the ontology of everyday artefacts was also addressed by more recent work in HCI~\cite{pradhan}. While these works shared the idea of applying a philosophical perspective with our work, they took a strictly ontological approach. In contrast, we focus on understanding the experience of technology, borrowing from HCI's tradition of using phenomenology. The concepts proposed in this paper address, due to their philosophical embedding, the perceived experience of objects and thus disregard the `objective' nature of the object. Consequently, technologies can be companions irrespective of their \emph{designed} social features. Thus, our work is applicable to inanimate artefacts and not specifically to social robots (which is the focus of HRI research).

Finally, we critically reflect on the limitations of our approach to conceptualising empathetic user experience. We should note that we did take a strong stance in constructing the framework around works in the philosophy of empathy. While we chose the works that form our framework with utmost care, the framework is dependent on assuming that certain philosophical stances are valid. The primary roles of the framework are serving as an analytical lens and providing a starting point for comparison and discussion. As long as no predictive powers are ascribed to the framework, we believe it can fulfil its role and its conceptual criticism can enable further contributions.



\section{Conclusion}
To broaden HCI's discussion on technologies as social agents, this paper outlined a framework for understanding the experience of companion technologies---interactive artefacts which can evoke empathy. We postulate that companion technologies have the potential to fulfil psychological needs left unaddressed by past generations frameworks of user experience. Our framework is built using notions from the philosophy of empathy, focusing on empathy for objects. We introduce four concepts that characterise empathetic experience: minded, feeling-experience, reflective and social significance. We showed that our framework concepts can be effectively used to analyse past work that focused on companion technologies to understand underlying psychological needs. We also proposed how our framework can be used with related concepts to better understand the experience of specific interactive artefacts.

We hope that our framework and the future research challenges for companion technologies provided can spark engaging discussions in the HCI community, support designers in exploring companion technologies in more depth and mitigate some of the potential challenges when designing technologies as companions.

\begin{acks}
We acknowledge the support of the Leibniz ScienceCampus Bremen Digital Public Health (lsc-diph.de), which is jointly funded by the Leibniz Association (W4/2018), the Federal State of Bremen and the Leibniz Institute for Prevention Research and Epidemiology---BIPS.
\end{acks}

\bibliographystyle{ACM-Reference-Format}
\bibliography{sample}

\end{document}